# Body Maps for Feeling of Creativity in Musical Chord Progression

Tatsuya Daikoku[1,2], Masaki Tanaka[1]

[1] Graduate School of Information Science and Technology, The University of Tokyo, Tokyo, Japan
[2] Centre for Neuroscience in Education, University of Cambridge, Cambridge, UK

**\* Corresponding author**
Tatsuya Daikoku
Graduate School of Information Science and Technology, The University of Tokyo, Tokyo, Japan,
7-3-1 Hongo, Bunkyo-ku, 113-8656, Tokyo, Japan
Email: daikoku.tatsuya@mail.u-tokyo.ac.jp

# Abstract

Creativity is a fundamental aspect of the human experience, enabling individuals to generate innovative ideas and artistic expressions including music. Although creativity is frequently framed as the ability to produce novel information, the mechanisms by which perceptions of creativity arise from this information—and their subsequent effects on our minds and bodies—remain inadequately understood.

This study investigates how musical stimuli evoke feelings of creativity through predictive mechanisms, while also examining the role of interoceptive bodily sensations, particularly those linked to the heart and stomach regions. Furthermore, we investigate the relationship between interoceptive sensibility and the intensity of the feeling of creativity.

By employing body-mapping assessments and emotional evaluations on 353 participants exposed to various chord progressions, we revealed the critical role of heart sensations in eliciting feelings of creativity. Our findings indicate that musical chord progressions characterized by high uncertainty and surprise generated heightened feelings of creativity alongside increased arousal. Notably, heart sensations correlated positively with the feeling of creativity, suggesting the crucial role of interoceptive bodily sensations in the experience of creativity. In contrast, the feelings of beauty and valence were more strongly influenced by predictable, low-uncertainty progressions, suggesting that the feelings of creativity may operate through a somewhat different cognitive mechanism than that of beauty and valence.

This study highlights the complex interplay between bodily sensations and feelings of creativity based on predictive processing. By framing creativity within the framework of interoception, our research contributes to a deeper understanding of the emotional and embodied processes that shape the experience of creativity. Future research should explore these relationships to enrich our comprehension of creativity in both artistic endeavors and everyday life.

# 1. Introduction

Creativity is an inherent aspect of human nature, defined as the ability to generate novel and valuable information, ideas, and arts including music (Lubart and Mouchiroud, 2003; Kozbelt et al., 2010; Robert, 2011). Throughout history, the question of how creativity arises in the brain has captivated many researchers. While it is widely recognized that creativity is deeply intertwined with accumulated knowledge, the precise processes underlying this relationship remain poorly understood. Notably, our understanding of how the feeling of creativity arises from acquired knowledge, and why certain information is perceived as creative, is still limited.

Importantly, creativity is often understood as the ability to produce novel information, encompassing art and ideas. However, a historical perspective shows that even great works of art and innovative ideas, including those in music, have not always been readily accepted by society. This suggests that the evaluation of seemingly identical novel information can vary depending on the mental and emotional states of the perceivers, including the creators themselves. When we consider the relationship between production and perception, we find that novelty resides in the information itself, while its creative value depends on the emotional responses evoked during perception. Thus, even when the information remains the same, whether it is regarded as creative is influenced by the sensibility of those experiencing it, including the creators.

One promising approach to unraveling this enigma focuses on the unique bodily sensations experienced during the perception of information (Putkinen et al., 2023; Hove et al., 2020). For instance, the feeling of creativity induced by music often evokes profound bodily sensations, such as a racing heart or a shiver down the spine. This suggests that, when listening to music, we engage not only our external sensory perceptions, such as those mediated by the auditory system (i.e., exteroception), but also our interoceptive sensation such as the acceleration of our heartbeat or spine-tingling thrills, and proprioception such as a feeling of constriction in the chest (Trost et al., 2017; Mori et al., 2017; Koelsch and Jancke, 2015). Consequently, our bodily sensations are intricately linked to our emotions and feelings.

Previous research has demonstrated that different emotions can be identified through the mapping of emotion-triggered bodily sensations (Nummenmaa et al., 2014). These studies indicate that various emotions have distinct bodily topographies, with negative emotions—such as fear, anger, sadness, and anxiety—activating the upper regions of the body, while positive emotions—such as happiness and love—engage a broader range of bodily responses. This implies that emotions are embodied, manifesting spatially within the body. Additionally, another study revealed that individual differences in interoceptive accuracy are associated with emotion-triggered bodily sensations (Jung, Ryu et al., 2017). This suggests that awareness of one's internal bodily states may serve as a critical messenger of sensory information during the emotional process. However, there is a notable lack of research on the relationship between the feeling of creativity and bodily sensations. Nonetheless, it is evident that feeling of creativity often provoke distinct bodily sensations, such as heightened excitement or exhilaration. Therefore, it appears that there is a close connection between feeling of creativity and bodily sensations.

The connection between bodily sensations and emotions can be better understood through the lens of the brain's predictive processing framework. The brain constructs emotions by minimizing the discrepancies between expected signals generated by its internal model and the sensory information received from exteroceptive, interoceptive, and proprioceptive sensations (Seth et al., 2013). It has been proposed that interoception, or the perception of internal bodily states, emerges from the ongoing refinement of these prediction errors (Barrett et al., 2015; Khalsa et al., 2015; Ainley et al., 2016). For example, when a musical chord progression concludes abruptly with an unexpected modulation to a different key, this can result in a sudden increase in interoceptive prediction error, such as an accentuated heartbeat. This heightened prediction error may then be addressed by adjusting interoceptive priors, potentially facilitated through deep breathing or other self-regulatory techniques.

In a recent study (Daikoku, Tanaka, Yamawaki, 2024), the authors developed a predictive model that mathematically decodes the relationship between perceptual uncertainty and surprise in music. Analyzing 80,000 chords from Billboard's US pop songs, they grounded their research in the framework of Western tonal harmony, interpreting it as a representation of musical syntax. They investigated emotional responses and the distribution of bodily sensations (embodied mappings) elicited by musical chords generated from this model. The results demonstrated that musical chords characterized by low uncertainty and high surprise elicited the greatest feelings of pleasure, with sensations predominantly localized to the heart region on the bodily map. Furthermore, when a preceding sequence of chords was associated with prediction errors, the resulting pleasure diminished. In contrast, sequences that were easy to predict enhanced feelings of pleasure and beauty, along with interoceptive sensations related to the heart. These findings imply that specific fluctuations in prediction may facilitate both interoceptive sensations linked to the heart and an enhanced musical sense of beauty. However, the relationship between the experience of musical creativity and bodily sensations within the framework of predictive processing remains largely unexplored and poorly understood.

This study embarks on a new exploration into how musical chord progressions relate to the feeling of creativity, using the body map of emotions as our guiding framework. Our primary goal is to determine whether certain types of musical chords, analyzed through the concept of predictive processing, can evoke feelings of creativity and whether these feelings are tied to interoceptive sensations in the body, particularly in the heart and stomach regions. These areas are known to play crucial roles in interoceptive awareness and prediction errors (Critchley and Harrison, 2013; Seth, 2013; Garfinkel et al., 2015). Additionally, we aim to investigate whether the strength of feeling of creativity is associated with interoceptive sensibility and to explore other emotional

factors—such as valence, arousal, and beauty—that may significantly influence the nature of these feeling of creativity. We hypothesize that, in addition to probabilistic properties, the interoceptive and bodily sensations induced by musical chord progressions and individual differences of interoceptive sensibility play a crucial role in eliciting feelings of creativity. That is, the feelings of creativity may be shaped by a combination of cognitive and embodied processes rather than musical predictability alone. Further, we postulate that other emotional factors—such as valence, arousal, and beauty influence the experience of the creativity.

# 2. Methods

## 2.1. Participants

This study involved body-mapping assessments followed by emotional evaluations across eight distinct 4-chord progressions (for details, see Daikoku et al., 2024). A total of 353 Japanese participants were included (mean age ± SD = 34.6 ± 9.96, 186 females, musical training experience ± SD = 3.1 ± 6.78 years). None of the participants had neurological or audiological conditions or absolute pitch. The experiment adhered to the Declaration of Helsinki and received approval from The University of Tokyo's Ethics Committee (Approval No. UT IST RE 230601). All participants provided informed consent and completed the tasks via personal computer.

## 2.2. Materials

The experimental paradigm was created using the Gorilla Experiment Builder (https://gorilla.sc), a cloud-based platform for conducting behavioral research online. Each participant was presented with eight different types of 4-chord progressions (500 ms per chord, 44.1 kHz, 32-bit, using Electric Piano 1 from General MIDI). The amplitude of the chords was adjusted based on equal-loudness-level contours to ensure consistent loudness perception across all participants. We employed a statistical-learning model (Daikoku, Minaotya et al., 2023) to estimate the surprise and uncertainty of each chord, calculating Shannon information content and entropy based on transitional probabilities (Shannon, 1948) from the McGill Billboard Corpus of 890 pop songs (Burgoyne, Wild, and Fujinaga, 2011). This implicit learning process allows the brain to compute transitional probabilities, assess uncertainty (Hasson, 2017), and predict future states using internalized models. Transitional probability ($P(e_{n+1}|e_n)$), derived from Bayes' theorem ($P(e_{n+1}|e_n)$), describes how the brain anticipates upcoming events, expecting higher-probability outcomes while being surprised by less likely events. The transitional probabilities are often translated as information contents:

$$I(e_{n+1}) = -\log_2 P(e_n+1|e_n)$$

Lower information content (i.e., higher transitional probability) corresponds to greater predictability and less surprise, while higher information content (i.e., lower transitional probability) signifies lower predictability and greater surprise. Consequently, a tone with lower information content is more likely to be predicted and selected by a composer as the next event compared to a tone with higher information content. In computational music studies, information content is essential for exploring prediction and statistical learning. The entropy of a chord $e_{n+1}$ represents the expected information content of chord $e_n$, calculated by summing the products of conditional probabilities and information contents of all possible chords in S:

$$H(e_{n+1}) = - \Sigma \, p(e_{n+1} = e|e_n)\log_2 p(e_{n+1} = e|e_n)$$

Entropy measures the listener's uncertainty when predicting an upcoming chord based on previous ones, while information content reflects the level of surprise upon hearing the chord. Using this framework, we designed 92 distinct chord progressions, categorized into eight types, which are available online (https://osf.io/cyqhd/). Each type features varying levels of uncertainty and surprise. Four types begin with low uncertainty and surprise (sLuL: a-d in Figure 1), and the others with high (sHuH: e-h in Figure 1). The fourth chord in each progression was varied using a 2x2 pattern of high/low uncertainty and surprise, with values based on the top and bottom 20% of the data. There are four possible variations for the fourth chord: the first has both low uncertainty and surprise (sLuL: Figure 1a and 1e), the second has low uncertainty but high surprise (sHuH: Figure 1b and 1f), the third shows high uncertainty with low surprise (sLuH: Figure 1c and 1g), and the fourth has both high uncertainty and surprise (sHuH: Figure 1d and 1h).

In the end, they consist of 1) sLuL-sLuL sequence (Figure 1a) representing the condition where the 1st-3rd chords have low surprise and uncertainty and the 4th chord has low surprise and uncertainty, 2) sLuL-sHuL sequence (Figure 1b) representing the condition where the 1st-3rd chords have low surprise and uncertainty and the 4th chord has high surprise and low uncertainty, 3) sLuL-sLuH sequence (Figure 1c) representing the condition where the 1st-3rd chords have low surprise and uncertainty and the 4th chord has low surprise and high uncertainty, 4) sLuL-sHuH sequence (Figure 1d) representing the condition where the 1st-3rd chords have low surprise and uncertainty and the 4th chord has high surprise and uncertainty, 5) sHuH-sLuL sequence (Figure 1e) representing the condition where the 1st-3rd chords have high surprise and uncertainty and the 4th chord has low surprise and uncertainty, 6) sHuH-sHuL sequence (Figure 1f) representing the condition where the 1st-3rd chords have high surprise and uncertainty and the 4th chord has high surprise and low uncertainty, 7) sHuH-sLuH sequence (Figure 1g) representing the condition where the 1st-3rd chords have high surprise and uncertainty and the 4th chord has low surprise and high uncertainty, and 8) sHuH-sHuH sequence (Figure 1h) representing the condition where the 1st-3rd chords have high surprise and uncertainty and the 4th chord has high surprise and uncertainty.

The thresholds for high and low values were determined from the top and bottom 20% of the data. Multiple chord progressions were generated for each of the eight types. Participants were randomly assigned progressions.

## 2.3. Procedures

All participants ($N$ = 353) conducted body-mapping tests and the following emotional judgements in every eight types of 4-chord progression. That is, the participants were presented with the eight types of chord progressions in a randomized sequence. After each session, they indicated where in their bodies they felt sensations by clicking on a body image displayed on the screen. Participants could make up to 100 clicks, and clicking was allowed both during and after listening to the sound (see Supplementary Figure S1 for further details). Participants rated their impression of each chord progression using a nine-point Likert scale. That is, after listening, they evaluated two emotional dimensions: valence and arousal. Additionally, they assessed how they felt about creativity and beauty from the musical chords using the same nine-point Likert scale. The scale ranged from 1 to 9, with 5 representing a neutral midpoint for both measures, allowing for a balanced assessment of each type of chord progression.

## 2.4. Statistical Analysis

We analyzed the body mapping data by extracting the total number of clicks in two key interoceptive regions: the cardiac and abdominal areas and another important area of head (Daikoku et al., 2024) for each participant, based on their x and y coordinate data. The raw x and y coordinates (see Supplementary Figure S2) were down-sampled by a factor of 40. To assess data normality, we applied the Shapiro-Wilk test to the total click counts in the cardiac and abdominal regions, as well as to the emotional ratings for valence, arousal, and feelings of creativity and beauty from the nine-point Likert scale assessments (see Supplementary Figure S3). Based on the normality test results, either parametric or non-parametric (Friedman) repeated-measures ANOVA was used to compare emotional ratings across the eight chord progression types. Emotional ratings for valence, arousal, and feelings of creativity and beauty served as dependent variables, with the chord progression types as the within-subject factor. Additionally, parametric or non-parametric (Spearman) correlation analyses were performed to examine relationships between click counts at cardiac, abdomen, and head positions and emotional scores. Statistical tests were conducted using jamovi (Version 1.2), with significance set at $p < .05$ and FDR used for multiple comparisons. Body topographies were generated in Matlab (2022b) by interpolating x and y coordinates with a meshgrid and color map.

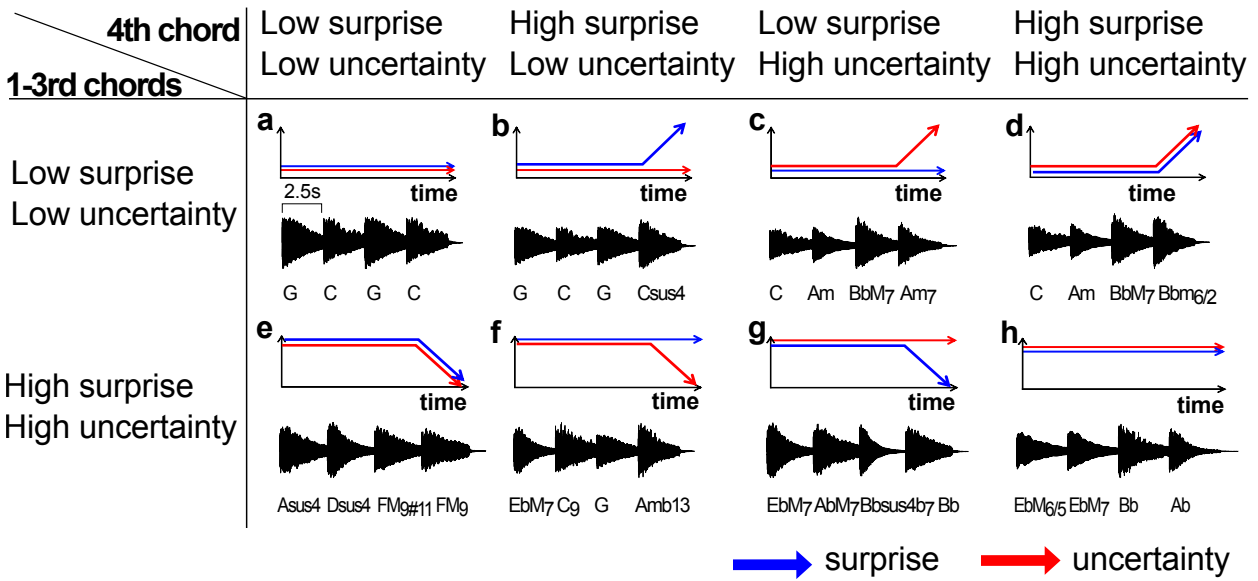

**Figure 1. Examples of chord progressions used in the study.** Reprinted from Daikoku et al., (2024). The eight types of chord sequences were generated using a statistical-learning model (Daikoku, Minaotya et al., 2023), which calculated Shannon information content and entropy. These metrics were derived from the transitional probabilities of each chord, based on a dataset of 890 pop songs from the US Billboard charts (Burgoyne, Wild, and Fujinaga, 2011). Entropy reflects a listener's uncertainty in predicting an upcoming chord (red lines), while information content measures the surprise upon hearing it (blue lines). The eight chord progressions vary in these factors. Four types start with low surprise and uncertainty (sLuL: a-d), and the other four begin with high surprise and uncertainty (sHuH: e-h). The fourth chord in each progression follows a 2x2 pattern, showing different combinations of high and low surprise and uncertainty. The first pattern exhibited both low surprise and low uncertainty (sLuL: a and e), the second had low uncertainty but high surprise (sHuL: b and f), the third showcased low surprise with high uncertainty (sLuH: c and g), and the fourth possessed both high surprise and uncertainty (sHuH: d and h). In the end, they consist of 1) sLuL-sLuL sequence (a) representing the condition where the 1st-3rd chords have low surprise and uncertainty and the 4th chord has low surprise and uncertainty, 2) sLuL-sHuL sequence (b) representing the condition where the 1st-3rd chords have low surprise and uncertainty and the 4th chord has high surprise and low uncertainty, 3) sLuL-sLuH sequence (c) representing the condition where the 1st-3rd chords have low surprise and uncertainty and the 4th chord has low surprise and high uncertainty, 4) sLuL-sHuH sequence (d) representing the condition where the 1st-3rd chords have low surprise and uncertainty and the 4th chord has high surprise and uncertainty, 5) sHuH-sLuL sequence (e) representing the condition where the 1st-3rd chords have high surprise and uncertainty and the 4th chord has low surprise and uncertainty, 6) sHuH-sHuL sequence (f) representing the condition where the 1st-3rd chords have high surprise

and uncertainty and the 4th chord has high surprise and low uncertainty, 7) sHuH-sLuH sequence (g) representing the condition where the 1st-3rd chords have high surprise and uncertainty and the 4th chord has low surprise and high uncertainty, and 8) sHuH-sHuH sequence (h) representing the condition where the 1st-3rd chords have high surprise and uncertainty and the 4th chord has high surprise and uncertainty.

# 3. Results

## 3.1. Emotion in response to musical chord progressions

All statistical analysis results, along with descriptive data for valence, arousal, and feelings of beauty and creativity from the nine-point Likert scale, have been made publicly available (https://osf.io/nw572/). The Shapiro–Wilk test indicated significant deviations from normality for valence, arousal, and beauty ratings ($p < .001$). Consequently, we used Friedman's non-parametric repeated-measures ANOVA to analyze the within-subject factor of the eight chord progression types.

The ANOVA for creativity revealed significant main effects ($\chi^2 = 20$, $p = 0.006$, Creativity). Post hoc analysis using the Durbin-Conover test indicated that creativity scores were significantly higher for chord progressions with consistently high uncertainty: sHuH-sLuH and sHuH-sHuH progressions (Figure 2g, 2h, Creativity) compared to the predictable chord progressions with low uncertainty: sLuL-sLuL progression (Figure 2a, Creativity) (sHuH-sLuH: $p = 0.042$, sHuH-sHuH: $p = .028$, Creativity). Additionally, creativity was significantly greater in sHuH-sLuH progression (Figure 2g, Creativity) compared to sLuL-sHuL progression (Figure 2b, $p = 0.042$, Creativity) and sLuL-sHuH progressions (Figure 2d, $p = 0.037$, Creativity). In summary, creativity tends to increase when chord progressions maintain high levels of uncertainty and prediction error. However, the strength of this effect is statistically small.

In this regard, the effect of arousal elicited by the prediction of chord progression is similar to the findings for creativity. Specifically, arousal levels were shown to increase in response to chord progressions with high prediction error and uncertainty, indicating that participants experienced heightened arousal when exposed to more unpredictable musical sequences as seen in creativity. This suggests that both creativity and arousal are significantly influenced by the high uncertainty and surprise in the musical chords. That is, the ANOVA for arousal revealed significant main effects ($\chi^2 = 154$, $p < 0.001$, Arousal). Post hoc analysis using the Durbin-Conover test indicated that arousal scores were significantly higher for chord progressions transitioning from unpredictable high-uncertainty chords to the last chord characterized by high uncertainty but low surprise: sHuH-sLuH progression (Figure 2g, Arousal) compared to the other progressions (all: $p < 0.002$). Additionally, arousal scores were significantly higher for the sHuH-sHuH progression (Figure 2h, Arousal) compared to the sLuL-sLuL, sLuL-sHuL, sLuL-sHuH, and sHuH-sHuH progressions (all: $p < 0.001$). The arousal scores were also significantly higher for the sHuH-sHuL progression (Figure 2f, Arousal) compared to the sLuL-sLuL (Figure 2a), sLuL-sHuL (Figure 2b), sLuL-sHuH (Figure 2d), sHuH-sHuH progressions (Figure 2h) (all: $p < 0.006$). On the other hand, arousal scores were also significantly higher for the sLuL-sLuH progression (Figure 2c, Arousal) compared to the sLuL-sLuL (Figure 2a), sLuL-sHuL (Figure 2b), sLuL-sHuH (Figure 2d), and sHuH-sHuH progressions (Figure 2h) (all: $p < 0.001$). The arousal scores were significantly lower for the sLuL-sHuH progression (Figure 2d, Arousal) compared to the sLuL-sHuL (Figure 2b, p = 0.017) and sHuH-sHuH progressions (Figure 2h, p = 0.00187). The arousal scores were significantly lower for the sHuH-sHuH progressions (Figure 2h) compared to the sHuH-sLuL (Figure 2e, p = 0.0016).

While the feelings of creativity and arousal are heightened by high levels of uncertainty and surprise in chord progressions, the feelings of beauty and valence are more strongly influenced by low uncertainty and low surprise. That is, the ANOVA for the feeling of beauty revealed significant main effects ($\chi^2$ = 70.3, p < 0.001). Post hoc analysis using the Durbin-Conover test indicated that beauty scores were significantly higher for chord progressions with consistently low uncertainty and low surprise: sLuL-sLuL progression (Figure 2a, Beauty) compared to the other progressions: sLuL-sHuL (Figure 2b, p = 0.014), sLuL-sLuH (Figure 2c, p = 0.028), sLuL-sHuH (Figure 2d, p = 0.009), sHuH-sLuL (Figure 2e, p = 0.014), sHuH-sHuL (Figure 2f, p = 0.009), sHuH-sLuH (Figure 2g, p = 0.007), and sHuH-sHuH progressions (Figure 2h, p = 0.006). Additionally, the beauty scores were significantly higher for sLuL-sHuL progression (Figure 2b) compared to the sHuH-sHuL (Figure 2f, p = 0.04), sHuH-sLuH (Figure 2g, p = 0.003), and sHuH-sHuH progressions (Figure 2h, p = 0.008). The beauty scores were significantly higher for sLuL-sHuH progression (Figure 2d) compared to the sLuL-sLuH (Figure 2c, p = 0.004), sHuH-sHuL (Figure 2f, p = 0.039), sHuH-sLuH (Figure 2g, p = 0.003), and sHuH-sHuH progressions (Figure 2h, p = 0.007).

The ANOVA for valence revealed significant main effects ($\chi^2$ = 108, p < 0.001). Post hoc analysis using the Durbin-Conover test indicated that valences were significantly higher for chord progressions with consistently low uncertainty and low surprise: sLuL-sLuL progression (Figure 2a, Valence) compared to the other progressions: sLuL-sHuL (Figure 2b, p = 0.002), sLuL-sLuH (Figure 2c, p < 0.001), sLuL-sHuH (Figure 2d, p = 0.011), sHuH-sLuL (Figure 2e, p < 0.001), sHuH-sHuL (Figure 2f, p = 0.009), sHuH-sLuH (Figure 2g, p = 0.007), and sHuH-sHuH progressions (Figure 2h, p = 0.006). Additionally, the valences were significantly higher for sLuL-sHuL progression (Figure 2b) compared to the sLuL-sLuH (Figure 2c, p = 0.004), sHuH-sLuL (Figure 2e, p = 0.004), sHuH-sHuL (Figure 2f, p = 0.004), sHuH-sLuH (Figure 2g, p = 0.003), and sHuH-sHuH progressions (Figure 2h, p = 0.010). The valences were significantly higher for the sLuL-sHuH progression (Figure 2d) compared to the sLuL-sLuH (Figure 2c, p = 0.003), sHuH-sLuL (Figure 2e, p = 0.003), sHuH-sHuL (Figure 2f, p = 0.002), sHuH-sLuH (Figure 2g, p = 0.002), and sHuH-sHuH progressions (Figure 2h, p = 0.002).

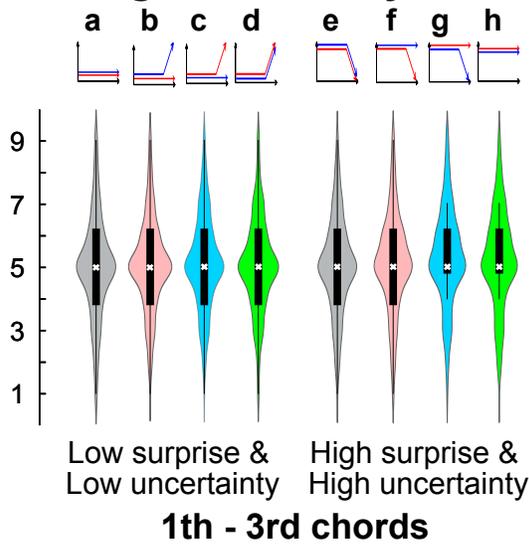

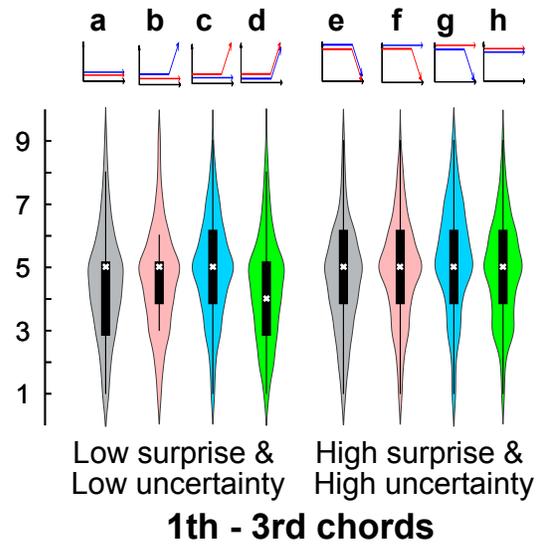

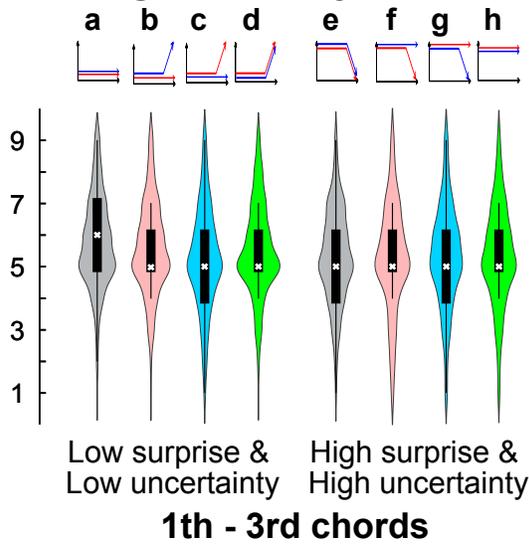

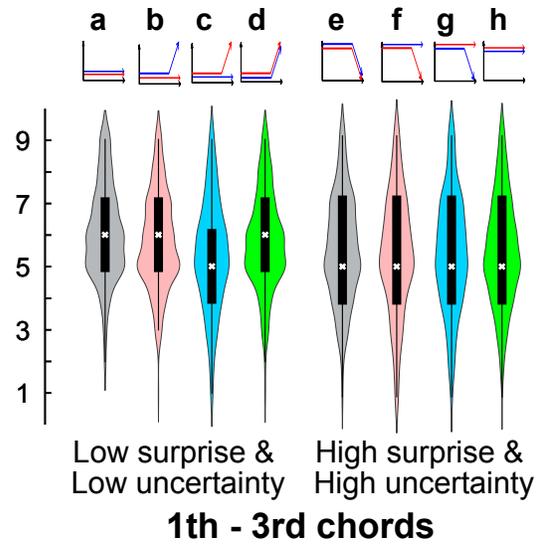

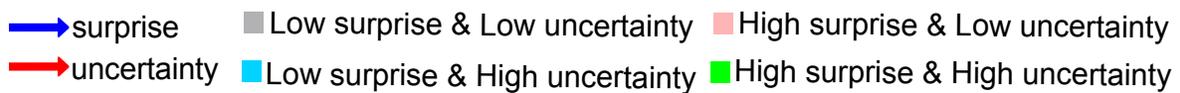

**Figure 2. Creativity, beauty, valence, and arousal ratings for musical chord progressions.**
The blue and red arrows represent the surprise and uncertainty values, respectively.

## 3.2. Bodily sensation and the emotions

While the statistical analyses indicate that the feeling of creativity is influenced by the predictability and uncertainty of musical chord progressions, this effect is notably weaker compared to other emotional responses such as valence, arousal, and the feeling of beauty. We hypothesize that, in addition to probabilistic properties, the interoceptive and bodily sensations induced by musical chord progressions and individual differences of interoceptive sensibility play a crucial role in eliciting feelings of creativity. That is, the feelings of creativity may be shaped by a combination of cognitive and embodied processes rather than musical predictability alone.

To test this hypothesis, we examined how bodily perception influences emotional experience: feeling of creativity and beauty as well as valence and arousal. We divided two groups including individuals with bodily sensation and without bodily sensation in each interoceptive area of the heart and abdomen. We also included another area of the head because the participants frequently showed the head sensations during hearing musical chord progressions (see Figure 3). Given the large sample size, which included all eight types of chord progressions (353 samples per chord type, totalling 2,824 samples), as the Central Limit Theorem indicates, deviations from normality have minimal impact in large samples, justifying the use of ANOVA. Therefore, we applied ANOVA to analyze the dependent variables of feeling of creativity and beauty, and valence and arousal. The fixed factors were bodily perception (i.e., with vs. without bodily perception) and the eight types of chord progressions, analyzed across different interoceptive regions of the heart and abdomen.

The ANOVA for the feeling of creativity revealed significant main effects of the heart (F = 6.554, p = 0.011, $\eta^2 p$ = 0.002) but not the abdomen (F = 2.214, p = 0.137, $\eta^2 p$ = 0.001) and head (F = 0.184, p = 0.668, $\eta^2 p$ = 0). This indicated that the feeling of creativity was significantly higher in individuals with heart sensations than in individuals without heart sensations (t = 2.56, p = 0.011, cohen's d = 0.118, 95%CI = 0.0275 – 0.208).

The ANOVA for the feeling of beauty revealed significant main effects of the heart (F = 6.513, p = 0.011, $\eta^2 p$ = 0.00) and the abdomen (F = 3.978, p < .001, $\eta^2 p$ = 0.01) but not head (F = 0.0178, p = 0.894, $\eta^2 p$ = 0). This indicated that the feeling of beauty was significantly higher in individuals with interoceptive sensations at the heart and abdomen, as compared with individuals without interoceptive sensations (t = 2.55, p = 0.011, cohen's d = 0.117, 95%CI = 0.0272 – 0.208).

The ANOVA for the valence revealed significant main effects of the heart (F = 13.442, p < .001, $\eta^2 p$ = 0.005) but not the abdomen (F = 0.205, p = 0.65, $\eta^2 p$ = 0) and head (F = 3.201, p = 0.074, $\eta^2 p$ = 0.001). This indicated that the valence was significantly higher in individuals with heart sensations than in individuals without heart sensations (p < .001).

The ANOVA for the arousal revealed significant main effects of the heart ($F = 7.113$, $p = 0.008$, $\eta^2 p = 0.003$) but not the abdomen ($F = 0.804$, $p = 0.37$, $\eta^2 p = 0$) and head ($F = 0.176$, $p = 0.675$, $\eta^2 p = 0$). This indicated that the arousal was significantly higher in individuals with heart sensations than in individuals without heart sensations.

Although the ANOVA for the feeling of creativity did not find the main effects of types of chord progressions ($p = 0.886$), the ANOVA for the feeling of beauty, valence, and arousal found the main effects of types of chord progressions (all: $p < .001$, see an external source for all results: https://osf.io/nw572/). These are consistent with the results of the ANOVA for the emotional responses (Figure 2). That is, the predictability and uncertainty of musical chord progressions does not strongly influence the feeling of creativity. Rather, the ANOVA suggests that the interoceptive and bodily sensations may influence the feelings of creativity. In summary, the feelings of creativity may be shaped by a combination of cognitive and embodied processes rather than musical predictability alone.

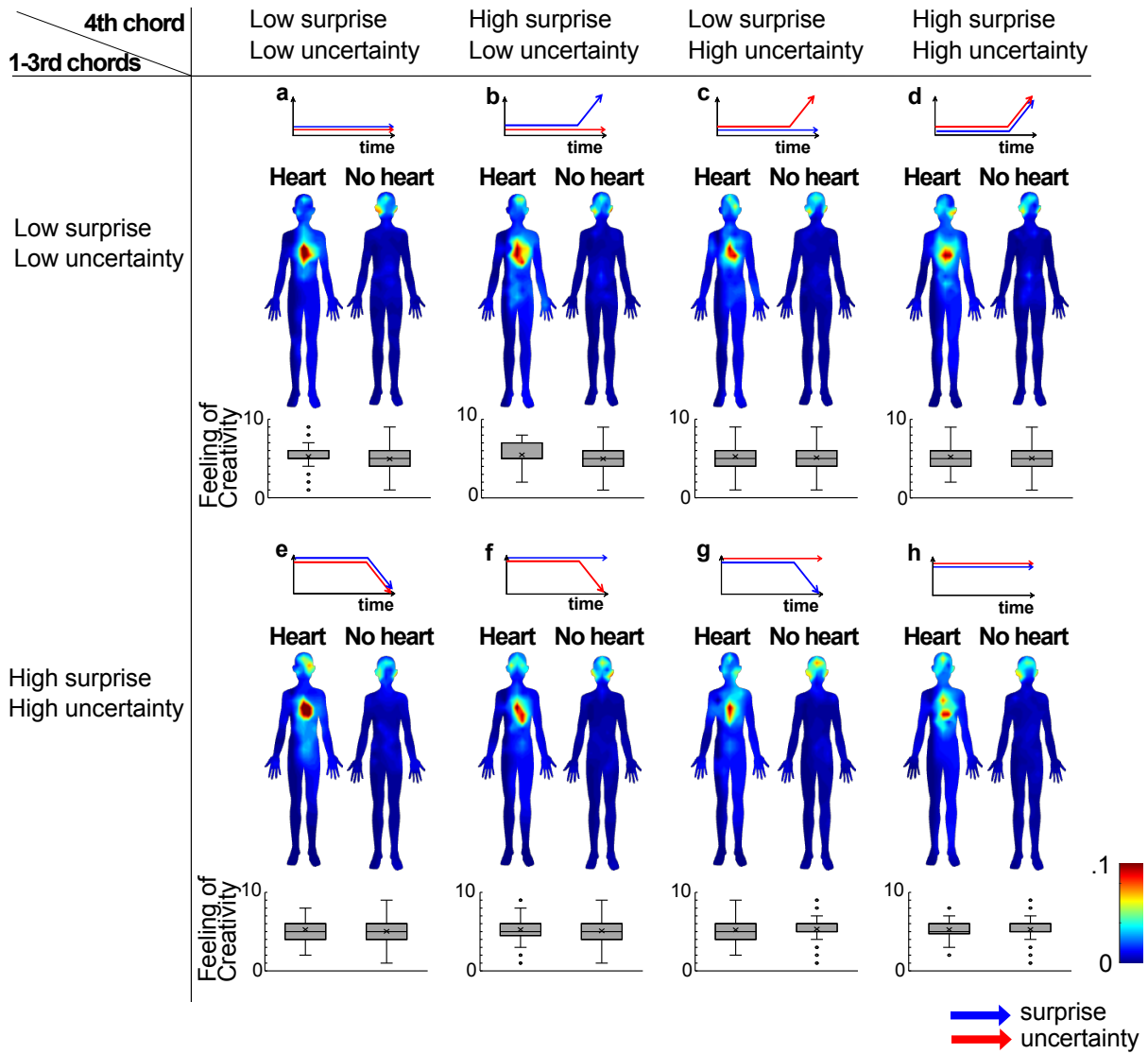

**Figure 3. Body topography of musical chord progressions in individuals with vs. without heart sensations.** The blue-to-red gradients represent the number of clicks. The blue and red arrows represent the surprise and uncertainty values, respectively. Heart = individuals with heart sensations during hearing musical chord progression. No heart = individuals without heart sensations during hearing musical chord progression.

## 3.3. Correlation between bodily sensation and the creativity

We also examined how the total number of clicks at cardiac and abdomen positions and the scores of body perception questionnaires (BPQ) evaluating interoceptive sensibility was correlated with the feelings of creativity in each type of chord progression. Further, we investigated how the feelings of creativity are correlated with the different emotional responses of valence, arousal, and beauty. The Shapiro–Wilk test for normality showed the violations of the assumption of normality on all the data ($p < .001$). Hence, we applied the Spearman correlation tests. All the results of statistical analyses and the descriptive have been deposited to an external source (https://osf.io/nw572/).

Results revealed significant positive correlations of creative feelings with the number of clicks localized to the cardiac area only in the sLuL-sHuL progressions transitioning from predictable chords to those characterized by low uncertainty and high surprise (Figure 4b) ($rs = 0.134, p = 0.024$). Additionally, we found significant positive correlations of creative feelings with the number of clicks localized to the abdomen area only in the sLuL-sHuH progressions transitioning from predictable chords to those characterized by high uncertainty and high surprise (Figure 4d) ($rs = 0.146, p = 0.013$). These suggest that interoceptive sensations at cardiac and abdomen areas are the important factors for the feelings of creativity in these types of chord progressions. Regarding the individual differences of interoceptive sensibility, we detected significant positive correlations of creative feelings with the score of BPQ in the sLuL-sHuH progressions transitioning from predictable chords to those characterized by high uncertainty and high surprise (Figure 5d, $rs = 0.14, p = 0.017$) and the sHuH-sLuH progressions transitioning from unpredictable chords to those characterized by high uncertainty and low surprise (Figure 5g) ($rs = 0.137, p = 0.021$). These suggest that individual interoceptive sensibility is the important factor for the feelings of creativity in these types of chord progressions. Finally, the feelings of creativity are positively correlated with the other emotional responses of valence, arousal, and beauty in each eight type of chord progression (all: $p < 0.001$).

**Figure 4. Correlation of the feeling of creativity with bodily sensations.** The blue-to-red gradients represent the number of clicks. The blue and red arrows represent the surprise and uncertainty values, respectively.

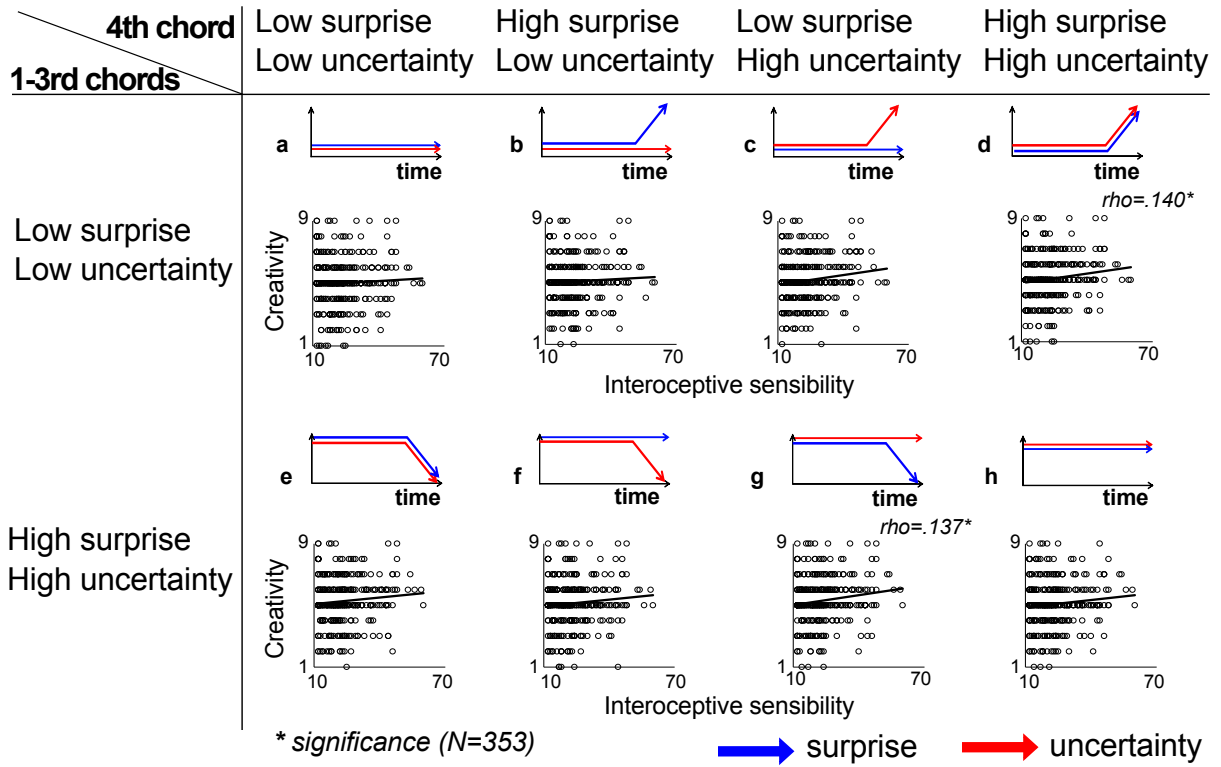

**Figure 5. Correlation of the feeling of creativity with interoceptive sensibility.** The blue-to-red gradients represent the number of clicks. The blue and red arrows represent the surprise and uncertainty values, respectively.

# 4. Discussion

The present study investigated which types of musical chord progressions evoke feelings of creativity through predictive processing, and whether these feelings are linked to interoceptive bodily sensations, particularly in the heart and abdomen regions. Additionally, we explored whether individual differences in interoceptive sensibility modulate the feeling of creativity. The results provide valuable insights into the role of bodily sensations in eliciting the feeling of creativity. Our findings demonstrate that the interoceptive sensations, particularly those associated with the heart and abdominal regions contribute to the feeling of creativity. This aligns with our hypothesis that creativity arises not solely from perceptual processes but also from embodied experiences, highlighting the importance of interoceptive awareness in fostering the feeling of creativity.

The results also indicate that musical chords characterized by high uncertainty and surprise tend to elicit stronger feelings of creativity and heightened arousal. This finding implies that both creativity and arousal are significantly influenced by high levels of uncertainty and surprise. In contrast, the feeling of beauty and valence were enhanced by low-uncertainty and predictable chord progressions. These results are in line with the findings in previous research (Daikoku et al., 2024), which detected evidence that the emotional responses of beauty and valence were stronger in the low-uncertainty and predictable chord progressions. These imply that, unpredictable stimuli, which are partially interpreted as subjectively new stimuli, can enhance the feeling of creativity and arousal, while predictable and low-uncertain stimuli may induce the feeling of valence and beauty. However, this study also found a positive correlation between creativity and beauty. The feelings of creativity elicited by unpredictable stimuli may interact with a unique sense of beauty that emerges in response to these unpredictable experiences, distinguishing it from the beauty associated with more predictable stimuli.

Additionally, we observed a positive correlation between heart sensations and feelings of creativity during the sLuL-sHuL progression, where the first three chords exhibit low surprise and uncertainty, while the fourth chord introduces high surprise with low uncertainty. This finding underscores the significant role of bodily sensations in enhancing the experience of creativity. Previous research also indicated that heart sensations were strongest during the sLuL-sHuL progression (Daikoku et al., 2024). It is possible that specific types of temporal dynamics of musical prediction particularly evoke interoceptive heart sensations, thereby triggering distinct emotional responses such as the feelings of creativity. A prior study also discovered that chords defined by a combination of low uncertainty and high surprise, as well as those with high uncertainty and low surprise, elicited the highest levels of pleasure (Cheung et al., 2019). This indicates that musical pleasure adheres to a two-dimensional inverted U-shaped curve characterized by the axes of uncertainty and surprise. Neurologically, this phenomenon is associated with activity in brain regions including the amygdala, hippocampus, and auditory cortex. Importantly, dopaminergic pathways in the nucleus accumbens, which demonstrate heightened responses to uncertainty, seem to play a crucial role in this effect. These findings align with Berlyne's seminal model, which asserts that pleasure exhibits an inverted U-shaped relationship with factors such as complexity. However, music exists within broader contexts that cannot be adequately captured by simply analyzing the predictions and uncertainties associated with individual chords. The surrounding context can significantly affect the perception of identical two-chord sequences, resulting in diverse emotional experiences. Consequently, this study emphasizes the need to expand our understanding of musical prediction and emotion through a three-dimensional model that incorporates preceding contexts, thereby facilitating a deeper exploration of temporal dynamics of musical prediction.

This study reveals that the positive correlation between creativity and interoceptive sensibility underscores the significance of individual differences in the experience of creativity. Individuals with heightened interoceptive sensibility may be more attuned to the bodily sensations elicited by musical stimuli, thereby facilitating a richer emotional experience and enhanced feelings of creativity. These findings pave the way for future research, suggesting that investigating interoceptive sensibility as a variable could provide deeper insights into the embodied cognitive processes that underpin the experience of creativity. Our previous studies indicated that individuals with low Body Perception Questionnaire (BPQ) scores (Tanaka, Daikoku, Yamawaki, 2024) and those exhibiting weak interoceptive awareness—such as individuals with alexithymia and depression (Daikoku, Horii, Yamawaki, 2023, preprint)—displayed less localized and more diffuse bodily sensations in response to musical chord progressions and pitched sounds, respectively. These diffuse bodily sensations are correlated with heightened feelings of anxiety and negative valence (Daikoku, Horii, Yamawaki, 2023, preprint), suggesting that such sensations may evoke negative emotions, including anxiety. Thus, emotional experiences induced by auditory perception are believed to involve the "embodiment" of sound through interoceptive pathways.

Despite the significant contributions of this study, several limitations warrant attention. First, the reliance on self-reported measures of creativity and emotional responses may introduce biases, as individual interpretations of creativity are inherently subjective and can vary widely among participants. Second, participants from a specific cultural background (i.e., Only Japanese) may limit the generalizability of the results to broader populations across different cultures. Future research could benefit from including a more diverse cohort to explore how cultural differences influence the relationship between interoceptive sensibility, bodily sensations, and feelings of creativity. Third, while this study focused on the heart and abdomen regions, other bodily areas that may contribute to the experience of creativity were not examined. Future investigations should consider a more comprehensive approach that includes various proprioceptive as well as interoceptive sensations, thereby providing a more nuanced understanding of how different bodily sensations interact with embodied cognitive processes related to feeling of creativity. Moreover, the study's design, which employed chord progressions as the primary stimulus, may not capture the full spectrum of musical experiences that can evoke creativity. Exploring different musical genres, styles, and contexts could yield valuable insights into the complex dynamics of how music influences creativity.

In conclusion, our findings highlight the crucial role of interoceptive bodily sensations in the experience of creativity. By framing creativity within the context of interoception, this study contributes to a deeper understanding of how embodied experiences shape creative expression. Further investigation into these roles will be essential for advancing our understanding of creativity in both artistic and everyday contexts.

## Competing Interests

The authors declare no competing financial interests.

## Author Contributions

T.D. and M.T. conceived the experimental paradigm and method of data analysis. T.D. collected the data, analysed the data, and wrote the draft of the manuscript and figure. T.D. and M.T. edited and finalized the manuscript.

## Acknowledgements

This research was supported by the Japan Science and Technology Agency (JST) Moonshot Goal 9 (JPMJMS2297), Japan. The funding sources had no role in the decision to publish or prepare the manuscript.

## Data Availability

All of anonymized raw data files, stimuli used in this study and the results of statistical analysis have been deposited to an external source (https://osf.io/nw572/). The other data are shown in supplementary data.